\documentclass[12pt]{article}

\usepackage{epsfig}
\usepackage{cite}
\usepackage{amsmath, amssymb, amsfonts}
\usepackage{color}
\usepackage{latexsym}  
\usepackage{graphicx}
\usepackage{cancel}
\usepackage[colorlinks,bookmarks]{hyperref}
\hypersetup{pdfpagemode=UseNone, pdfstartview=FitH, linkcolor=blue, citecolor=red, urlcolor=blue}

\def\be{\begin{equation}}
\def\ee{\end{equation}}
\def\ba{\begin{eqnarray}}
\def\ea{\end{eqnarray}}

\def\bdm{\begin{displaymath}}
\def\edm{\end{displaymath}}
\def\la{~\mbox{\raisebox{-.6ex}{$\stackrel{<}{\sim}$}}~}
\def\ga{~\mbox{\raisebox{-.6ex}{$\stackrel{>}{\sim}$}}~}
\def\bq{\begin{quote}}
\def\eq{\end{quote}}

 at 10truept

\newcommand{\p}{\partial}

\renewcommand{\[}{\left[}
\renewcommand{\]}{\right]}

\newcommand{\Mpl}{M_{\mathrm{Pl}}}

\newcommand{\bea}{\begin{eqnarray}}
\newcommand{\eea}{\end{eqnarray}}

\newcommand{\bi}{\begin{itemize}}
\newcommand{\ei}{\end{itemize}}

\newcommand{\beq}{\begin{equation}}
\newcommand{\eeq}{\end{equation}}
\newcommand{\beqa}{\begin{eqnarray}}
\newcommand{\eeqa}{\end{eqnarray}}
\newcommand{\mpl}{\Mpl}

\def\la{~\mbox{\raisebox{-.6ex}{$\stackrel{<}{\sim}$}}~}
\def\ga{~\mbox{\raisebox{-.6ex}{$\stackrel{>}{\sim}$}}~}

\def\12{{1 \over 2}}

\def\ltap{\ \raise.3ex\hbox{$<$\kern-.75em\lower1ex\hbox{$\sim$}}\ }
\def\gtap{\ \raise.3ex\hbox{$>$\kern-.75em\lower1ex\hbox{$\sim$}}\ }
\def\gl{\ \raise.5ex\hbox{$>$}\kern-.8em\lower.5ex\hbox{$<$}\ }
\def\roughly#1{\raise.3ex\hbox{$#1$\kern-.75em\lower1ex\hbox{$\sim$}}}

\begin{document}

\thispagestyle{empty}
\begin{flushright}
June 2025 
\end{flushright}
\vspace*{1.5cm}
\begin{center}

 {\Large \bf Discretely Evanescent Dark Energy}
 
\vspace*{1.3cm} {\large
Nemanja Kaloper\footnote{\tt
kaloper@physics.ucdavis.edu} }\\
\vspace{.3cm}
{\em QMAP, Department of Physics, University of
California, Davis, CA 95616, USA}\\

\vspace{1.5cm} ABSTRACT
\end{center}
We propose a new UV-complete dark energy model which
is \underbar{\it neither} a cosmological constant nor
a slowly rolling scalar field. Our dark energy is the flux
of a top form in a hidden sector gauge theory similar to QCD.
The top form controls the vacuum energy generated by dark sector CP violation.
Its flux discharges by the nucleation of membranes that source it.
The tension and charge of the membranes are set
by the chiral symmetry breaking scale $\sim 10^{-3} eV$, and the dark
energy is a transient. It decays on the order of the current age of the
universe. The decays decrease dark energy discretely and randomly,
instead of gradually like rolling scalars.
Since the decay rate is close to the present Hubble
scale, $\Gamma \ga H_0^4$, 
in a time $\sim {\cal O}(1/H_0)$ the cosmic acceleration may even cease altogether.

\vfill \setcounter{page}{0} \setcounter{footnote}{0}

\vspace{1cm}
\newpage

\section{Introduction}

Cosmological observations show that about $70\%$ of the critical energy density of our Universe today
is dark energy (most recently, \cite{DESI:2024mwx}). This dark energy component is commonly modeled
as a cosmological constant \cite{Weinberg:1987dv,Martel:1997vi} 
or quintessence: a slowly varying scalar field with a very shallow potential \cite{Wetterich:1994bg,Zlatev:1998tr}. 
Adopting either of these models of dark energy, one either ignores the elephant in the room -- 
the cosmological constant problem -- or assumes that it is relegated to the domain of 
anthropics (see \cite{Weinberg:1988cp,Polchinski:2006gy}
for a review). 

In the case of slowly varying scalar fields the problem is typically further aggravated. 
The scalar field potential must be very shallow, with the initial field displacement from the vacuum very large. 
The field theoretic challenges led to some effective field theories for quintessence, for example 
based on axion-like fields or their avatars 
\cite{Frieman:1995pm,Fukugita:1994hq,Nomura:1999py,Armendariz-Picon:2000nqq,Armendariz-Picon:2000ulo,Kim:2002tq,Kaloper:2008qs,DAmico:2018mnx}. 
However there are also foundational challenges to field theory. 
Many dark energy models yield universes whose expansion accelerates forever, which have  
cosmological event horizons \cite{Hellerman:2001yi,Fischler:2001yj}. It is still unclear 
how to formulate unitary quantum field theory from
first principles in such spacetimes, or whether that is even possible. 

All these problems may be alleviated in models of dynamical dark energy which 
disappears after some time. In those cases, the dark energy equation of state parameter $w = p/\rho$ 
may be detectably different from $-1$. Curiously, some new observations 
also \cite{DESI:2024mwx} suggest that dark energy might be changing in time. 
As tentative as this may be at the moment, 
it is interesting to consider first-principles models that can naturally 
yield such dynamics. After presenting our model
and its cosmological evolution, we also discuss observational implications.

\section{Framework}

Following recent work on addressing the strong CP problem using the discrete relaxation
of the vacuum energy in the CP-violating vacua \cite{Kaloper:2025wgn,Kaloper:2025upu}, we propose a 
UV complete model of dark energy which arises naturally from chiral symmetry breaking in a hidden sector
which consists of a $SU(N)$ gauge theory with quark-like fermions, defined by a perturbative Lagrangian
\be
{\cal L}_{\tt dark} = - \frac{1}{4} G^a_{\mu\nu} G^{a~\mu\nu} 
+ \sum_{l,m} Tr\Big( \bar \psi_l \bigl(i \delta_{lm} \cancel{D} - M_{lm}\bigr) \psi_m \Bigr) 
+ \frac{g^2 \theta_{\tt dark}}{64\pi^2} \epsilon^{\mu\nu\lambda\sigma} \sum_a G^a_{\mu\nu} G^a_{\lambda\sigma} \, .
\label{darkl}
\ee 
Here $G^a_{\mu\nu}$ are the components of the dark gauge field strength of the non-Abelian Yang-Mills 
vector potential $B^a_{\mu}$, given by 
$G^a_{\mu\nu} = \partial_\mu B^a_{\nu} - \partial_\nu B^a_{\mu} - g f^{abc} B^b_{\mu} B^c_{\nu} $, and
$f^{abc}$ are the non=Abelian structure constants.
The fermions $\psi_l$ are the dark quarks, whose flavor is counted by $l$. The matrix 
$M_{lm}$ is the dark mass matrix. The tracing is over the color index, 
treating $\psi$ as a color fundamental $\psi = \{ \psi^\alpha \}$ and $\bar \psi$ as its conjugate, and defining
$i \cancel D = i \gamma^\mu \partial_\mu \delta^{\alpha\beta} - \frac{g}{2} \sum_a B^a_\mu (\tau^a)^{\alpha\beta}$,
where $\frac12 \tau^a$ are the gauge group generators, and  
$M_{lm} = M_{lm} \delta^{\alpha\beta}$. We will treat $M$ as a subleading contribution to the dark quark masses, 
which is technically natural since when $M \rightarrow 0$ the theory is chiral. 

We include the vacuum angle
term $\propto \theta_{\tt dark}$ anticipating the role it will play further on in contributing to, and controlling dark energy
evolution. The physical vacuum angle $\hat \theta$ is then 
composed of the sum of $\theta_{\tt dark}$ and the overall phase of the
dark quark mass matrix ${\tt Arg} \det { M}$. To make this manifest, we change the fermion
basis using chiral transformations, so that in the basis where $\det {M}$
is real, the total CP-violating phase multiplying $Tr(G {}^*G)$ in the perturbative Lagrangian includes
the phase of the original basis $\det {M}$,
\be
\theta_{\tt dark} \rightarrow  \hat \theta = \theta_{\tt dark} + {\tt Arg} \det {M} \, .
\label{cptheta}
\ee

For simplicity, we assume that one of the dark quarks is exactly massless 
before chiral symmetry breaking, so that  the 
determinant of ${M}$ is zero during that time. Since in that case the phase ${\tt Arg} \det {M}$ is 
completely arbitrary, the total CP-violating phase 
$\hat \theta$  can be chosen to be zero. Therefore there is no physical manifestation of CP violation
before dark chiral symmetry is broken. The theory splits into completely degenerate superselection
sectors parameterized by $\hat \theta$. 

Since we are merely interested in addressing non-constant dark energy, we ignore the 
`big' cosmological constant problem \cite{Weinberg:1988cp}, and assume it is solved 
independently of late dark energy.
As jarring as this might be, it is a common practice and we shall adopt it here. We will be agnostic as to how
the problem is solved, and simply assume that in the early
universe before dark chiral symmetry breaking dark energy is zero. 
This logic may be affirmed if the observations like \cite{DESI:2024mwx} eventually 
conform that $w \ne -1$.

Next, let us imagine that forces of nature unify near 
$M_{\tt GUT} \sim 10^{16} GeV$. This is not necessary for the model but it reduces the phase space
for model building in a convenient way, by setting a specific boundary condition for the couplings in the 
theory, simplifying the picture of the UV completion. To see this we use the one-loop 
renormalized coupling constant 
with a $SU(N)$ gauge group and
$n_f$ fermions, given by 
\be
\alpha_{\tt dark}(\mu) = \frac{g^2}{4\pi} = \frac{6\pi}{\Bigl|11N - 2 n_f \Bigr| \ln(\mu/\Lambda_{\tt dark})} \, .
\label{running}
\ee
We can then readily check that at the GUT scale, 
$\alpha_{\tt dark}(M_{\tt GUT}) \simeq 0.027/(N -2n_f/11)$. 
For it to be comparable to e.g. either the electroweak or the gravitational coupling, 
$\alpha_{\tt dark}(M_{\tt GUT}) \sim 
\alpha_{\tt EW} \sim \alpha_{\tt grav} \sim M_{\tt GUT}/\mpl \sim 10^{-2}$. 
Thus $N \sim {\cal O}({\rm few})$ and $n_f < {\cal O}(10)$ works  
when the chiral symmetry breaking occurs at a very low scale, 
$\Lambda \sim 10^{-3} eV$, as indicated by the fact that $\alpha_{\tt dark} \gg 1$ at scales $\mu < 10^{-3} eV$. 

\section{Universe Before Dark Chiral Symmetry Breaking}

In the early universe, before dark chiral symmetry breaking and dark confinement,
practically all the dark quarks and gluons 
evolve like a mostly uniform relativistic plasma in equilibrium for a long time, almost until the present epoch, 
but decoupled from the visible sector due to only interacting with it gravitationally. The dark temperature $T_{\tt dark}$ 
of this sector is set by dark reheating at the end of inflation. This temperature would 
initially be \underbar{at least} 
\be
T_{\tt dark} \sim H_* \, , 
\label{darkst}
\ee
where $H_*$ is the
Hubble parameter at the end of inflation. This follows from universality of gravity, and specifically is 
the result of gravitational particle production, which is universal \cite{Parker:1968mv,Bunch:1978yq,Linde:1990flp}. 
The temperature 
could be higher if either the inflaton directly reheats the dark sector, or the visible sector reheating is
very inefficient. Ultimately, the firm criterion for the dark sector temperature comes from BBN, 
which is that the dark sector radiation energy density can be at most about a percent of the visible one.
This means that $T_{\tt dark}$ could be as high as ${\cal O}(0.3) \, T_{\tt CMB}$, but not more.
Thus the dark sector must be colder than the visible sector. Note that even if this extreme bound were saturated, with 
 $T_{\tt dark}/T_{\tt CMB} \sim 0.3$, the late universe cosmological bounds on the relativistic species 
 $\Delta N_{\tt eff}$ are still satisfied  \cite{Berezhiani:2000gw,DAmico:2017lqj}.
  
With efficient visible sector reheating, $T_{\tt visible}$ can be as high as 
$T_{\tt visible} \sim \sqrt{\mpl H_*}$. 
Using the bounds on the ratio of tensor and scalar primordial density fluctuations\footnote{Gravitational particle production of dark sector 
particles yielding (\ref{darkst}) would also induce some isocurvature perturbations, but those are very small compared to the adiabatic inflationary 
perturbations.} \cite{BICEP:2021xfz},
$r \la 0.036$, and the fact that 
\be
r \sim \Bigl|\frac{\delta_{\tt T}}{\delta_{\tt S}}\Bigr|^2 \sim 10^{10} \Bigl(\frac{H_*}{\mpl} \Bigr)^2 \, ,
\label{r}
\ee
this 
means that in this case initially 
\be
\frac{T_{\tt dark}}{T_{\tt visible}} \la \sqrt{\bigl(10^{-5} \sqrt{r}\bigr)} \sim 10^{-3} \, .
\label{ts}
\ee
This ratio is preserved through the adiabatic stages of cosmic expansion. 
However the freezeout of some visible sector species will amplify this ratio. Every time a visible sector particle decouples 
as the visible temperature drops below its mass, the total visible entropy, and hence effective visible temperature,
drop. This could enhance the ratio by a factor of $10$ or so, maybe even $\la 100$  by now, cranking it up
to $0.01 - 0.1$ or so. 

Information about the dark temperature is important in order to
determine when the chiral symmetry breaking phase transition happens. 
If the dark sector were too hot, the chiral symmetry breaking 
would be delayed. However, from the reheating analysis above, 
if in the late universe $T_{\tt dark}/T_{\tt visible} \sim T_{\tt dark}/T_{\tt CMB} \la 0.01$, this means that when the
dark sector temperature is $T_{\tt dark} \sim 10^{-3} eV$, the CMB temperature is at at least few $eV$. Since the 
dark chiral symmetry breaking happens when $T_{\tt dark} \la \Lambda_{\tt dark}$, in our case
this will happen around, or before, recombination and radiation-matter equality, depending on the initial ratio
$T_{\tt dark}/T_{\tt visible}$. Bear in mind that this still allows dark chiral symmetry breaking significantly before recombination and radiation-matter equality. 

After the dark chiral symmetry breaking, most of the dark sector 
particles will have gained weight, with masses $\simeq 10^{-3} eV$. This follows from Eq. (\ref{running}) since
$\mu \sim \Lambda_{\tt dark}$ near the chiral symmetry breaking at 
$\alpha_{\tt dark} \sim 1$, and we imagine that 
the dark quarks get the mass solely, or predominantly, from chiral symmetry breaking. 
The dynamics being completely controlled by a single dimensional quantity $\Lambda_{\tt dark}$, 
other dimensional parameters which appear in perturbation theory are comparable to it. 
The early dark radiation would be converted in part to the cooling warm dark matter with $m \sim 10^{-3} eV$ 
and to dark energy. Some dark radiation might remain as lightest 
dark glueballs. A nice sketch of the relevant physics before and around this stage can be 
gleaned from \cite{Georgi:1984zwz}, using the discussion in
Section 3.2, ``Toy Model," after changing some numbers and 
suppressing quark masses arising outside of QCD. 

The dark chiral symmetry breaking is triggered by the formation of the dark chiral condensate,
breaking of the dark quark chiral symmetry and corresponding disappearance
of a null eigenvalue of the dark quark mass matrix. As a result, $\det M \ne 0$, and the 
dark CP-violating phase $\hat \theta$ is fixed to a definitive value. CP symmetry in the dark sector is broken,
and the condensate gives rise to the $\hat \theta$-dependent vacuum energy
$V_{\tt dark} = {{{\cal X}_{\tt dark}}} {\cal V}(\hat \theta)$.
The dimensional quantity ${\cal X}_{\tt dark}$ is the topological susceptibility of the theory, 
${\cal X}_{\tt dark} \simeq \Lambda_{\tt dark}^4$, and ${\cal V}$ is a function computed in the instanton expansion.

A simple approach to understand the dynamics of dark chiral symmetry breaking is to use the 
language of the gauge theory top form, which has been initiated by L\"uscher, who derived the 
it from the nonperturbative contributions to the correlation functions of the duals of gauge theory 
anomaly terms below chiral symmetry breaking \cite{Luscher:1978rn}. Although the top form does
not have local fluctuating degrees of freedom, it can change due to nontrivial topological configurations
in the theory, which for example can behave as charged membranes 
\cite{Kaloper:2025wgn,Kaloper:2025upu,Dvali:2005an,Dvali:2005zk}, quantum-mechanically relaxing the flux
of the top form. Since the top form locally is degenerate with a cosmological constant 
 \cite{Aurilia:1978qs,Aurilia:1980xj,Duff:1980qv,Aurilia:1980jz}, and its magnetic dual is degenerate with the 
 CP-violating phase $\hat \theta$, this opens up the channel for simultaneously 
 relaxing both the total CP-violating phase and $V_{\tt dark}$ toward zero.
 This follows since  $\theta_{\tt dark} = 0 + 2n\pi$, which is the only CP-invariant superselection sector, 
 is the minimum of the potential  $V_{\tt dark}$ \cite{Vafa:1984xg}.

\section{Dark {\it SU(N)} and the Emergence of Its Top Form}

Turning to the details of the phase with broken dark chiral symmetry, we focus on the 
anomaly term 
\be
\frac{g^2}{64\pi^2} \epsilon^{\mu\nu\lambda\sigma}  \sum_a G^a_{\mu\nu} G^a_{\lambda\sigma} = 
\frac{g^2}{16\pi^2}  \epsilon^{\mu\nu\lambda\sigma} \partial_\mu 
\Bigl( \sum_{a} B^a_\nu \partial_\lambda B^a_\sigma 
+ \frac{g}{3} \sum_{a,b,c} f_{abc} B^a_\nu B^b_\lambda B^c_\sigma \Bigr) 
= \frac{1}{2\pi^2} \partial_\mu K^\mu \, ,
\label{anomals}
\ee
where $f_{abc}$ are structure constants and $K_\mu$ is the Chern-Simons current. We can trade $K^\mu$ 
for its Hodge dual $3$-form, $K_\mu = \epsilon_{\mu\nu\lambda\sigma} A^{\nu\lambda\sigma}/6 $, which, since
\be
\partial_\mu K^\mu = \frac16 \epsilon^{\mu\nu\lambda\sigma} \partial_\mu A_{\nu\lambda\sigma} = 
\frac{1}{24} 
\epsilon^{\mu\nu\lambda\sigma} F_{\mu\nu\lambda\sigma} \, , 
~~~~ {\rm with} ~~~~ F_{\mu\nu\lambda\sigma}
= 4  \partial_{[\mu} A_{\nu\lambda\sigma]} \, , 
\label{forms}
\ee 
yields 
\be
F_{\mu\nu\lambda\sigma} = \frac34 g^2 \sum_a G^a_{[\mu\nu} G^a_{\lambda\sigma]} \, ,
\label{4formcpterm}
\ee
where $[\ldots]$ denotes antisymmetrization over the enclosed indices. 
Hence the anomaly term in the perturbative Lagrangian ${\cal L}_{\tt QCD}$ is  
\be
\frac{1}{48 \pi^2} \theta_{\tt dark} \epsilon_{\mu\nu\lambda\sigma} F^{\mu\nu\lambda\sigma} \, .
\label{4theta}
\ee
After dark chiral symmetry breaking, when the null eigenvector of the dark quark mass matrix $M$ is
lifted, so that $\det M \ne 0$, and after transforming to the dark quark basis where $\det(M)$ is real, 
we replace $ \theta_{\tt dark}$ in Eq. (\ref{4theta}) with $\hat \theta$ as per Eq. (\ref{cptheta}). 

Next we compute the nonperturbative actions to the theory below chiral symmetry breaking. Integrating out
the dark quarks first, we are interested in the expectation values of operators ${\cal O}$ which only depend 
on the gauge field variables \cite{Luscher:1978rn}. This yields  
\be
\langle {\cal O} \rangle_{\hat \theta} = \frac{1}{{\cal Z}[{\hat \theta}]} 
\int \[{\cal D}{\cal B}\] {\cal O} \, e^{i S + i \hat \theta \int q(x)} \, , ~~~~~~ 
{\cal Z}[{\hat \theta}] =  \int \[{\cal D}{\cal B}\] \, e^{i S + i \hat \theta \int q(x)} \, .
\label{expec}
\ee
${\cal Z}[{\hat \theta}]$ is the standard partition
function with the ``source" term $ \hat \theta \int q(x)$ and the operator $q(x)$ is 
$q(x) = \frac{g^2}{64\pi^2} \epsilon^{\mu\nu\lambda\sigma} 
\sum_a G^a_{\mu\nu}(x) G^a_{\lambda\sigma}(x)  = \frac{1}{48 \pi^2} \epsilon_{\mu\nu\lambda\sigma} 
F^{\mu\nu\lambda\sigma}$. $\langle \ldots \rangle_{\hat \theta}$  
represents averaging with the Euclidean path integral of the gauge theory including
the anomaly term $\propto \hat \theta$, and the integration is over the Yang-Mills vector potential
$B^a_\mu$.

The vacuum expectation value of ${\cal O} = q(x)$ is \cite{Luscher:1978rn}
\be
\langle q \rangle_{\hat \theta} = i f(\hat \theta) \, ,
\label{expq}
\ee
where by translational invariance, 
$\langle q(x) \rangle_{\hat \theta} = \langle q(0) \rangle_{\hat \theta} \propto 
i \partial_{\hat \theta}{\cal Z}[\hat \theta]/{\cal Z}[\hat 
\theta]$. Therefore, $f(\hat \theta)$ is an odd function, 
$f(-\hat \theta) = -f(\hat \theta)$  \cite{Luscher:1978rn}. 
Further, the $3$-form potential $A_{\mu\nu\lambda}$ gains the propagator. 
In Lorenz gauge $\partial_\mu A^{\mu\nu\lambda} = 0 $, the propagator is 
given by the connected Green's function computed with
the $\hat \theta$-term in the action:
\ba
&&~~\frac{1}{(3!)^2} \epsilon_{\mu\alpha\beta\gamma} \epsilon_{\nu\sigma\rho\delta}  
\langle A^{\alpha\beta\gamma}(x) A^{\sigma\rho\delta}(y)\rangle_{\hat \theta} 
= \int \frac{d^4p}{(2\pi)^4} e^{ip(x-y)} \frac{p_\mu p_\nu}{p^2} G_{\hat \theta}(p^2) \, , \nonumber \\
&& p^2 G_{\hat \theta}(p^2) = - 4\pi^4 i \int d^4x \, \langle T\bigl(q(x)q(0)\bigr) \rangle_{\hat \theta} = 
-4\pi^4 i \, \partial_{\hat \theta} \langle q(0) \rangle_{\hat \theta} = 4\pi^4  {\cal X_{\tt dark}}\, .
\label{greens}
\ea

The residue of the momentum space Green's function 
at $p^2 \rightarrow 0$ is the dark topological susceptibility 
${\cal X_{\tt dark}} \simeq (\Lambda_{\tt dark})^4$ of the theory, 
defined by the integral of the time-ordered correlator $T\bigl(q(x)q(0)\bigr)$: 
\be
{\cal X}_{\tt dark} =  - i \int d^4x \, \langle T\bigl(q(x)q(0)\bigr) \rangle_{\hat \theta} \, .
\label{suscept}
\ee
Its precise form is model-dependent. Some explicit examples 
are discussed by L\"uscher in \cite{Luscher:1978rn}.
Since the momentum space Green's function is
$p^2 G_{\hat \theta}(p^2) = -4\pi^4 i \, \partial_{\hat \theta} \langle q(0) \rangle_{\hat \theta} = 
4\pi^4 \partial_{\hat \theta} f({\hat \theta})$, the function 
$f$ is $f(\hat \theta) = {\cal X}_{\tt dark} \hat \theta + \ldots$.
Next, we use this to deduce the quadratic $4$-form 
field strength $F_{\mu\nu\lambda\sigma}$ in the action; its 
presence must reproduce the Green's function (\ref{greens}).
Rescaling the $3$-form potential by ${\cal A}_{\mu\nu\lambda} 
= \frac{1}{2\pi^2 \sqrt{\cal X}_{\tt dark}} \, A_{\mu\nu\lambda}$, 
a straightforward algebra  \cite{Kaloper:2025wgn,Kaloper:2025upu} shows 
the top form sector in the dark Lagrangian is 
\be
{\cal L}_{\tt dark} \ni \frac{\sqrt{{\cal X}_{\tt dark}}}{24} 
\hat \theta \epsilon_{\mu\nu\lambda\sigma} {\cal F}^{\mu\nu\lambda\sigma}  
- \frac{1}{48} {\cal F}_{\mu\nu\lambda\sigma}^2  \, .
\label{efflag} 
\ee
up to gauge-fixing terms. A related analysis for QCD 
was given in \cite{Gabadadze:1997kj,Gabadadze:2002ff} .

To calculate the condensate-induced potential which sets the vacuum energy of the CP-violating
superselection sectors, we trade the electric field strength ${\cal F}_{\mu\nu\lambda\sigma}$ for its magnetic dual
$ \sim {\cal F}_{\mu\nu\lambda\sigma} \epsilon^{\mu\nu\lambda\sigma}/4!$. We follow
\cite{Kaloper:2025wgn,Kaloper:2025upu} verbatim, 
recasting the top form action in the first order form, and integrating the electric
top form. The electric action with Lagrange multiplier is 
\be
S_{\cal F} = \int d^4 x \Bigl(- \frac{1}{48} {\cal F}_{\mu\nu\lambda\sigma}^2 
+ \frac{\sqrt{{\cal X}_{\tt dark}}}{24}  \hat \theta \epsilon_{\mu\nu\lambda\sigma} {\cal F}^{\mu\nu\lambda\sigma}  
+ \frac{\cal F}{24}  \epsilon^{\mu\nu\lambda\sigma} 
\bigl({\cal F}_{\mu\nu\lambda\sigma} - 4 \partial_\mu {\cal A}_{\nu\lambda\sigma} \bigr) \Bigr) \, ,
\label{cantra}
\ee
so that after shifting the
electric $4$-form variable 
$\tilde {\cal F}^{\mu\nu\lambda\sigma} = {\cal F}^{\mu\nu\lambda\sigma} 
- \bigl(\sqrt{{\cal X}_{\tt dark}} \hat \theta + {\cal F} \bigr) \epsilon^{\mu\nu\lambda\sigma}$
and integrating it out we obtain
\be
S_{\cal F} = \int d^4 x \Bigl(- \frac{{\cal X}_{\tt dark}}{2}  
\bigl(\hat \theta + \frac{\cal F}{\sqrt{\cal X}_{\tt dark}} \bigr)^2 - \frac{\cal F}{6}  
\epsilon^{\mu\nu\lambda\sigma}  \partial_\mu {\cal A}_{\nu\lambda\sigma} \Bigr) \, .
\label{cantrad}
\ee
Now the  
field equations are
\be
\delta {\cal A}_{\nu\lambda\sigma} : ~~~ \partial_\mu {\cal F} = 0 \, , 
~~~~~~~ \delta {\cal F}: ~~~ \epsilon^{\mu\nu\lambda\sigma} {\cal F}_{\mu\nu\lambda\sigma} 
= 4 \epsilon^{\mu\nu\lambda\sigma}\partial_{[\mu}{\cal A}_{\nu\lambda\sigma]} 
= - 4! \Bigl(\sqrt{{\cal X}_{\tt dark}}\hat \theta + {\cal F} \Bigr) \, ,
\label{eomsc}
\ee
and the potential $V_{\tt dark} = {\cal X}_{\tt dark} {\cal V}$, tuned to zero at the minimum and truncated to
the quadratic term, is 
\be
V_{\tt dark} = \frac{{{\cal X}_{\tt dark}}}{2}  \bigl(\hat \theta + \frac{\cal F}{\sqrt{{\cal X}_{\tt dark}}} \bigr)^2 \, .
\label{potential}
\ee
This is precisely the contribution to the vacuum energy when CP is violated (see 
e.g. \cite{Gabadadze:1997kj,Gabadadze:2002ff}). 
The total CP-violating phase in the dark sector is
\be
\theta_{\tt dark} = \hat \theta + \frac{\cal F}{\sqrt{{\cal X}_{\tt dark}}} \, .
\label{totphase}
\ee
The expression for $\theta_{\tt dark}$ becomes more intricate with the addition of new 
phases and nonperturbative corrections, which we ignored. These can all be systematically accounted for.

We can invert the Hodge dual in Eq. (\ref{eomsc}) to find 
\be
{\cal F}_{\mu\nu\lambda\sigma} = \bigl(\sqrt{{\cal X}_{\tt dark}} \hat \theta 
+ {\cal F} \bigr) \epsilon_{\mu\nu\lambda\sigma} 
= \sqrt{{\cal X}_{\tt dark}}  \theta_{\tt dark} \, \epsilon_{\mu\nu\lambda\sigma} \, ,
\label{soln4form}
\ee
as per Eq. (\ref{totphase}). This includes the totally arbitrary magnetic dual of 
top form, ${\cal F}$, which is an as-yet unspecified 
integration constant. We see that the vanishing of $\theta_{\tt dark}$ and ${\cal F}_{\mu\nu\lambda\sigma}$
is in one-to-one correspondence. This means, that 
CP restoration in the dark sector, by any means which can reduce the top form flux, is equivalent
to $\theta_{\tt dark}$ relaxation toward zero, which simultaneously reduces the 
dark energy $V_{\tt dark}$. In other words, once dark chiral symmetry breaks, dark energy will be induced. But as long
as there exist channels for discharging ${\cal F}_{\mu\nu\lambda\sigma}$, dark energy will not be forever. It
will decay and disappear after a time. 

It has long been argued that this happens because the theory contains gauge field 
configurations which are charged under the top form 
\cite{Gnadig:1976pn,Aurilia:1978dw,Luscher:1978rn,Aurilia:1980xj,Gabadadze:1997kj,Gabadadze:2002ff}. 
In QCD  below chiral symmetry breaking, the charged objects are membranes, 
the spherical domain walls separating the deconfined vacua inside and confining vacua outside. 
Our theory is a scaled-down QCD, and so the membranes arise in similar ways\footnote{An alternative may be to 
introduce a top form external to the dark gauge theory, which mixes with the dark anomalous current, and 
membranes charged under it controlled by a scale $\la \Lambda_{\tt dark}$. It is not obvious 
that in this nonminimal model vacuum energy would be eventually completely cancelled, as explained below.}. 
These charged walls maintain Gauss's law for the top form, 
and also balance the energy difference between the interior and exterior. 
Indeed, by recalling Eq. (\ref{efflag}) we can interpret the bilinear term as a boundary term,
$\frac{\sqrt{{\cal X}_{\tt dark}}}{24} \hat \theta \epsilon_{\mu\nu\lambda\sigma} {\cal F}^{\mu\nu\lambda\sigma} 
= \frac{\sqrt{{\cal X}_{\tt dark}}}{6} \hat \theta \epsilon_{\mu\nu\lambda\sigma} \partial^\mu {\cal A}^{\nu\lambda\sigma}$. 
Integrating this over the spherical surface bounding a region 
with ${\cal F}_{\mu\nu\lambda\sigma} \ne 0$ and ${\cal F}_{\mu\nu\lambda\sigma}=0$ yields
\be
S_{\cal F} = - \int d^4 x \frac{1}{48} {\cal F}_{\mu\nu\lambda\sigma}^2 + 
\frac {\sqrt{{\cal X}_{\tt dark}}\hat \theta}{6} \int d^3 \xi \, 
{\cal A}_{\mu\nu\lambda} \frac{\p x^\mu}{\p \xi^a} \frac{\p x^\nu}{\p \xi^b} 
\frac{\p x^\lambda}{\p \xi^c} \epsilon^{abc} \, ,
\label{thetacharge}
\ee
where the second term involves an integral over the sphere separating
deconfined interior and an exterior vacuum with all fields confined inside a small ball.

This is precisely the phenomenological charge term for a thin membrane, as seen in the papers 
\cite{Gabadadze:1997kj,Gabadadze:2002ff,Gnadig:1976pn,Aurilia:1978dw}, 
\be
{\cal Q} = - \sqrt{{\cal X}_{\tt dark}} \hat \theta \, .
\label{qcdcharge}
\ee
To complete the membrane Lagrangian, one would also
need to add the tension term, which must arise as a measure of the rest energy of a membrane,
essentially gapping it from the vacuum \cite{Gnadig:1976pn}, which has the form
\be
- {\cal T} \int d^3 \, \xi \sqrt{|\det(\eta_{\mu\nu} \frac{\p x^\mu}{\p \xi^a} \frac{\p x^\nu}{\p \xi^b} )|} \, \, ,
\label{tension}
\ee
where the integration is over the membrane worldvolume, and ${\cal T}$ is the membrane tension. 
From this perspective, one notes that QCD in the strong coupling regime below chiral symmetry breaking
``manufactures" it's own system of membranes, under the guise of the boundaries of glueballs. 
Qualitatively the dark sector gauge theory we are considering differs little from this. 
The main difference is the chiral symmetry breaking 
scale, which for real QCD is $\Lambda_{\tt QCD} \simeq 100 MeV$, while for us it is 
$\Lambda_{\tt dark} \simeq 10^{-3} eV$.

After we include the membranes, we expect by comparison with 
\cite{Aurilia:1978dw,Aurilia:1980xj}, but also 
\cite{Brown:1987dd,Brown:1988kg,Kaloper:2022oqv,Kaloper:2022utc}, that the magnetic dual
top form flux ${\cal F}$ is quantized in the units of charge ${\cal Q}$,
\be
{\cal F} = {\cal N} {\cal Q} = - {\cal N} \sqrt{{\cal X}_{\tt dark}} \hat \theta \, ,
\label{quantized}
\ee
where ${\cal N}$ is an integer. Then, the quantum nucleation of membranes 
\cite{Brown:1987dd,Brown:1988kg,Kaloper:2022oqv,Kaloper:2022utc}, completely analogous to the 
Schwinger processes for particle production \cite{Schwinger:1951nm}, yields a discharge channel for ${\cal F}$.
That in turn would relax $\theta_{\tt dark}$, and the potential $V_{\tt dark}$, in discrete steps. When the
membrane charge is given by (\ref{qcdcharge}), and the flux is quantized as in (\ref{quantized}), 
these processes will relax the phase $\theta_{\tt dark}$ and $V_{\tt dark}$ exactly to zero. Since
\be
\Delta \theta_{\tt dark} = \frac{\Delta {\cal F}}{\sqrt{{\cal X}_{\tt dark}}} =  \hat \theta \, ,
\label{totphasechange}
 \ee
inside the bubble the phase changes by a unit of $\hat \theta$, and the potential $V_{\tt dark}$ by a corresponding 
shift. 

For a typical natural value of $\hat \theta \sim {\cal O}(1)$, and the natural initial condition
$\theta_{\tt dark} \sim {\cal O}(1)$ modulo $2\pi$, the corresponding region will relax after at most 
a few nucleations, as we will discuss below, since we will take charges of order $(\Lambda_{\tt dark})^2$.
If the initial value of the constant $\hat \theta$ is smaller, it may take more\footnote{This scenario may require 
a readjustment of the parameters that control the nucleation rates to make it fast enough for
final steps, if it is to completely cancel vacuum energy, as we examine below.}. However, since the 
effective total CP violating phase is $\theta_{\tt dark} = (1 - {\cal N}) \hat \theta$, the terminal value will
be a minimum of $V_{\tt dark}$. We will see it below shortly.   
We note that a similar behavior of the theory was introduced by Dvali in \cite{Dvali:2005zk}, who postulated that
the membrane charge depends on the top form flux. 

We have ignored this feature of the membranes ``native" to the gauge theory \cite{Kaloper:2025wgn,Kaloper:2025upu} 
out of concern that their nucleation rates could be highly suppressed, based on estimates of nucleation rates in
\cite{Shifman:1998if} and \cite{Forbes:2000et}. There are examples where these rates might not be as slow in the
large-N limit of a gauge theory \cite{Dvali:1998ms,Dubovsky:2011tu}. 

However such questions are of 
lesser concern to us here. In the case of real world QCD, the problem was ensuring that CP is restored by BBN, 
implying it needs to happen quickly after chiral symmetry breaking. In the case of dark energy, the only issue in the natural setup we are considering is if the ${\cal O}(1)$ factors allow, or prevent, slightly faster discharge rates. 
In a spacetime where the decay rates could be slightly faster future event horizons will
be absent, and vice versa. Of course, this will affect observational signatures and the prospects for detection. 
We will discuss this in more detail in what follows. 

\section{Quantal Evolution of Dark Energy}

Up until now, we were focused on explaining how a dark gauge theory can induce a contribution to the 
vacuum energy of the universe below dark chiral symmetry breaking. 
Let us now turn to the nucleation processes which can cancel this
vacuum energy at late times in more detail. Working below the dark chiral symmetry breaking scale, we can 
focus on the top form sector of the theory. 
Our action, including membranes with tension and charge, is 
\ba
S_{\cal F} &=& \int d^4 x \Bigl(- \frac{{\cal X}_{\tt dark}}{2}  
\bigl(\hat \theta + \frac{\cal F}{\sqrt{{\cal X}_{\tt dark}}} \bigr)^2 - \frac{\cal F}{6}  
\epsilon^{\mu\nu\lambda\sigma}  \partial_\mu {\cal A}_{\nu\lambda\sigma} \Bigr) \, .
\\
&-& {\cal T} \int d^3 \, \xi \sqrt{|\det(\eta_{\mu\nu} \frac{\p x^\mu}{\p \xi^a} \frac{\p x^\nu}{\p \xi^b} )|} 
- \frac{\cal Q}{6} \int d^3 \xi \, {\cal A}_{\mu\nu\lambda} \frac{\p x^\mu}{\p \xi^a} \frac{\p x^\nu}{\p \xi^b} 
\frac{\p x^\lambda}{\p \xi^c} \epsilon^{abc}  \, , \nonumber 
\label{cantradcharged}
\ea
using magnetic dual flux ${\cal F}$. 

We need the Euclidean action to calculate the membrane 
nucleation rates \cite{Coleman:1977py,Callan:1977pt,Garriga:1993fh}.
The reformulation of (\ref{cantradcharged}) to Euclidean space is straightforward; we follow the
steps in \cite{Kaloper:2022oqv,Kaloper:2022utc}, but we ignore the gravitational terms. 
First, we Wick-rotate the action using $t = - i x^0_E$, which gives 
$- i \int d^4x \sqrt{g} {\cal L}_{\tt QFT} = - \int d^4x_E \sqrt{g} {\cal L}^E_{{\tt QFT}}$. 
Next, using the conventions 
${\cal A}_{0 jk} = {\cal A}^{E}_{0jk}$, ${\cal A}_{jkl} =  {\cal A}^{E}_{jkl}$ and 
$\epsilon_{0ijk} = \epsilon^{E}_{0ijk}$ and $\epsilon^{0ijk} = -\epsilon_E^{0ijk}$, 
the tension and charge terms, with notation
$\gamma = |\det(\eta_{\mu\nu} \frac{\p x^\mu}{\p \xi^a} \frac{\p x^\nu}{\p \xi^b} )|$, transform to
$- i {\cal T} \int d^3 \xi \sqrt{\gamma} = - {\cal T} \int d^3 \xi_E \sqrt{\gamma}$ 
and $i {\cal Q} \int {\cal A}_i = - {\cal Q} \int {\cal A}_i$, and 
the scalars do not change. The Euclidean action, defined by $i S = - S_E$, after integrating the bilinear
term by parts to reflect boundary conditions on the membranes, is 
\ba
S_E&=&\int d^4x_E \Bigl( \frac{{\cal X}_{\tt dark}}{2}  \bigl(  \hat \theta+ \frac{\cal F}{\sqrt{{\cal X}_{\tt dark}}}   \bigr)^2 
+  \frac{1}{6} 
{\epsilon^{\mu\nu\lambda\sigma}_E} \partial_\mu \bigl( {\cal F} \bigr) {\cal A}^E_{\nu\lambda\sigma} \Bigr)  \nonumber \\
\label{actionnewmemeu}
&+& {\cal T} \int d^3 \xi_E \sqrt{\gamma}_{\cal A} - \frac{{\cal Q}}{6} \int d^3 \xi_E \, {\cal A}^E_{\mu\nu\lambda} \, 
\frac{\p x^\mu}{\p \xi^\alpha} \frac{\p x^\nu}{\p \xi^\beta} 
\frac{\p x^\lambda}{\p \xi^\gamma} \epsilon_E^{\alpha\beta\gamma} \, .
\ea
From here on we drop the index ${E}$. 

One more change of variables simplifies the analysis because 
$\hat \theta$ and ${\cal F}/\sqrt{{\cal X}_{\tt dark}}$ are completely degenerate in the actions above.
Further, since $\hat \theta$ is a constant after chiral symmetry breaking, all it does is shift the initial value of the
flux ${\cal F}$. Thus we can treat their sum $\theta_{\tt dark} = \hat \theta + {\cal F}/\sqrt{{\cal X}_{\tt dark}}$ 
as the new variable, defining the new $3$-form 
${\cal Y}_{\nu\lambda\sigma} = \sqrt{{\cal X}_{\tt dark}} {\cal C}_{\nu\lambda\sigma}$ to
maintain normalizations. In these variables, the new Euclidean action is 
\ba
S_{{\theta_{\tt dark}}} &=& \int d^4 x \Bigl(\frac{1}{2} {\cal X}_{\tt dark} \bigl(\theta_{\tt dark} \bigr)^2 
+ \frac{1}{6}  
\epsilon^{\mu\nu\lambda\sigma}  \partial_\mu \bigl( {\theta_{\tt dark}} \bigr) 
{\cal Y}_{\nu\lambda\sigma} 
\Bigr) \\
&-& {\cal T} \int d^3 \, \xi \sqrt{|\det(\eta_{\mu\nu} \frac{\p x^\mu}{\p \xi^a} \frac{\p x^\nu}{\p \xi^b} )|} 
- \frac{\cal Q}{6 \sqrt{{\cal X}_{\tt dark}}} \int d^3 \xi \, 
{\cal Y}_{\mu\nu\lambda} \frac{\p x^\mu}{\p \xi^a} \frac{\p x^\nu}{\p \xi^b} 
\frac{\p x^\lambda}{\p \xi^c} \epsilon^{abc}  \, . \nonumber 
\label{truncatedCP}
\ea
It is now obvious that the membrane discharges relax $\theta_{\tt dark}$ in finite
steps of $\frac{\cal Q}{\sqrt{{\cal X}_{\tt dark}}}$. 
Further it is also obvious that as ${\cal X}_{\tt dark} \rightarrow 0$,
and so chiral symmetry is restored, the discharge processes are inaccessible. The 
CP-violating  contribution to the potential vanishes in this limit, and the effective charge
$\frac{\cal Q}{\sqrt{{\cal X}_{{\tt dark}}}}$ for
$\theta_{\tt dark}$ diverges when ${\cal X}_{\tt dark} \rightarrow 0$ This makes the barrier to tunneling 
impassable in this limit, as we will see directly shortly.

From (\ref{truncatedCP}) we can construct the tunneling configurations, composed of slices of a 
$4D$ sphere $S^4$ glued along a fixed $S^3$ latitude sphere. These are
Euclidean worldvolumes of a spherical membrane with tension ${\cal T}$ and charge ${\cal Q}$. 
Detailed investigation of these configurations is given 
in \cite{Brown:1987dd,Brown:1988kg,Kaloper:2022oqv,Kaloper:2022utc}. In the 
limit $\mpl \rightarrow \infty$, only one of those 
configurations can be realized. It is the Euclidean bounce 
relating the backgrounds with non-negative vacuum energy 
\be
V_{\tt dark} = \frac{1}{2} {\cal X}_{\tt dark} \bigl(\theta_{\tt dark}\bigr)^2 \, ,
\label{finpots}
\ee
where the interior and exterior $\theta_{\tt dark}$ differ by a unit of charge, since 
$n^\mu \partial_\mu {\theta_{\tt dark}} =\frac{\cal Q}{\sqrt{{\cal X}_{\tt dark}}}\delta(r-r_0)$ 
\cite{Brown:1987dd,Brown:1988kg,Kaloper:2022oqv,Kaloper:2022utc}, 
and where $n^\mu$ is the outward-oriented normal to the membrane, $r_0$ its 
radius and $r$ the coordinate along $n^\mu$. 
Integrating across the wall gives
\be
\Delta {\theta_{\tt dark}} = \frac{\cal Q}{\sqrt{{\cal X}_{\tt dark}}} \, ,
\label{delH}
\ee
as claimed. Another constraint on membrane nucleation comes from energy conservation. 
The energy difference $\Delta V$ over the volume of the membrane interior, gained from
discharge, must compensate the energy cost for creating a membrane with tension ${\cal T}$ \cite{Coleman:1977py}. 
To balance the two, we look at the difference between the action  (\ref{truncatedCP}) for a configuration with 
one bubble and the action for a smooth initial background without bubbles. 
Since both $\Delta V$ and ${\cal T}$ are constant, the  
volume factors are $V_{S^4} = \pi^2 r_0^4/2$ and $V_{S^3} = 2\pi^2 r_0^3$, and the result is \cite{Coleman:1977py}
\be
S_{membrane} = 2\pi^2 r_0^3 {\cal T} - \frac12 \pi^2 r_0^4 \Delta V \, ,
\label{sbounce}
\ee
where $\Delta V= {\cal X}_{\tt dark} \theta_{\tt dark} \Delta \theta_{\tt dark}$. 
Minimizing (\ref{sbounce}) with respect to $r_0$ yields the 
membrane radius at nucleation \cite{Coleman:1977py},
\be
r_0 = \frac{3{\cal T}}{\Delta V } \, , 
\label{radnuc}
\ee
and so, substituting this into (\ref{sbounce}), yields the bounce action \cite{Coleman:1977py}
\be
B = \frac{27\pi^2}{2} \frac{{\cal T}^4}{\bigl(\Delta V\bigr)^3} \, .
\label{bounce}
\ee

The bubble nucleation rate per unit time per unit volume then is given by 
$\Gamma = A e^{-B}$ \cite{Coleman:1977py,Callan:1977pt}. To compute it we need
the prefactor $A$. Where gravity is negligible, the prefactor has been calculated by Garriga in 
\cite{Garriga:1993fh}. Using his results, we finally find
\be
\Gamma \simeq 9 \frac{{\cal T}^4}{\bigl(\Delta V\bigr)^2} 
\exp\Bigl({-\frac{27\pi^2}{2} \frac{{\cal T}^4}{\bigl(\Delta V\bigr)^3}}\Bigr) \, .
\label{nucrate}
\ee

The other tunneling configurations 
\cite{Brown:1987dd,Brown:1988kg,Kaloper:2022oqv,Kaloper:2022utc}
which exist when $\mpl < \infty$ are all much more suppressed than this flat space bounce
for the range of charges and tensions which we will use, especially when the background fluxes are
small. Since the discharges reduce the total flux and the 
dark energy density to exactly zero, those processes
are therefore irrelevant, and we will ignore them altogether. 
We will verify this shortly by noting that the
fast flat space bounce configurations, which we focus on, 
only utilize bubbles which have very small radius at nucleation. 

At this stage we can consider quantitative features of the model. As we noted in Sec. 2, the dark gauge theory
can easily accommodate the dark chiral symmetry breaking at $\Lambda_{\tt dark} \sim ({\cal X}_{\tt dark})^{1/4} \sim 
10^{-3} eV$. Further, as explained in Sec. 3, this will happen before the visible sector cools down to 
$T_{\tt CMB} \sim 10^{-3} eV$, i.e. this will have occurred in our past. Thus the dark dynamics will preset the
universe with a dark energy component with energy density 
\be
V_{\tt dark} = \frac{1}{2} {\cal X}_{\tt dark} \bigl(\theta_{\tt dark}\bigr)^2 \sim {\cal O}(1) \times (10^{-3} eV)^4 \, ,
\label{phende}
\ee
with typical and most likely values of $\theta_{\tt dark} \sim {\cal O}(1)$. There could be some
causally disconnected universes with $\theta_{\tt dark} \ll 1$, and correspondingly smaller
dark energy, but those are much less common. 
This is precisely as is required to fit our observations, and will initiate well before the epoch of low redshift $z \la 1$
where dark energy is currently probed. This dark energy is unstable to discharge of the top form, 
mediated by the nucleation of membranes charged under it. 
Thus to fit the observations, the dark energy must be sufficiently long lived -- which implies that it should
last at least as long as the current age of the universe, $1/H_0 \sim 10^{10}~{\rm years}$, where $H_0$
is the current Hubble parameter. Substituting 
$\Delta V \simeq {\cal X}_{\tt dark} { \theta_{\tt dark}} \Delta { \theta_{\tt dark}}
=  \sqrt{{\cal X}_{\tt dark}} \cal{Q} \, { \theta_{\tt dark}}$, into Eq. (\ref{nucrate}),
the nucleation rate per unit volume per unit time of membrane-bounded bubbles
which reduce the value of $\theta_{\tt dark}$ by a unit is 
\be
\Gamma \simeq 9 \frac{{\cal T}^4}{{\cal X}_{\tt dark} {\cal Q}^2 ( \theta_{\tt dark})^2 } 
\exp\Bigl({-\frac{27\pi^2}{2} \frac{{\cal T}^4}{({{\cal X}_{\tt dark}})^{3/2} \,{\cal Q}^3 ( \theta_{\tt dark})^3}}\Bigr) \, .
\label{nucratecp}
\ee
Recalling that ${\cal T} = \zeta (\Lambda_{\tt dark})^3$ and ${\cal Q} = \xi (\Lambda_{\tt dark})^2$, 
with $\zeta$ and $\xi$ being
${\cal O}(1)$, and that ${\cal X}_{\tt dark} \simeq (\Lambda_{\tt dark})^4$, 
the bounce action and the prefactor in Eq. (\ref{nucratecp}) are
\be
B = \frac{27\pi^2 \zeta^4}{2 \xi^3} \frac{1}{( \theta_{\tt dark})^3} \, , ~~~~~~~~~~~~~
A = \frac{9 \zeta^4}{ \xi^2} 
\Bigl({\Lambda_{\tt dark}}\Bigr)^4
\frac{1}{( \theta_{\tt dark})^2} \, .
\label{bth}
\ee

Requiring that the nucleation rates are fast enough to begin to efficiently 
discharge dark energy around now implies that 
\cite{Kaloper:2025wgn,Kaloper:2025upu,Guth:1982pn,Turner:1992tz}. 
\be
\Gamma \sim H_0^4 \, ,
\label{fastdec}
\ee
Using $\Lambda_{\tt dark} \sim 10^{-3} eV$, $\theta_{\tt dark} \sim 1$, and $H_0 \sim 10^{-33} eV$,
fixing $\xi = 1$ (i.e. ${\cal Q} = (\Lambda_{\tt dark})^2$),   
we finally obtain the numerical condition
\be
\zeta^4 \times e^{-133 \zeta^4} = \zeta^4 \times 10^{-58\zeta^4} \sim 10^{-121} \, ,
\label{numerika}
\ee
which shows that we need $\zeta^4 \sim 2$. In turn this implies that the tension should be
around ${\cal T} \simeq 1.2 (\Lambda_{\tt dark})^3$. Given e.g. the calculations of the glueball 
spectrum in QCD by \cite{Gabadadze:1997kj}, this appears not only achievable, but also natiral. 
If the tension is higher, this is 
not a problem since in that case the dark energy condensate would be more stable, delaying the
``thawing" of the cosmic permafrost. The only problem might occur if the tension is too small, in which case
the dark energy could dissipate too soon. However this requires $\zeta < 1$ and seems unnatural, 
to be ignored at a model-builder's discretion
at this moment. 

We can also confirm that in the limit $\Lambda_{\tt dark} \rightarrow 0$, where the 
dark susceptibility vanishes, the membrane discharge stops, as highlighted above. This can be seen from the vanishing
of the decay rate in Eq. (\ref{nucratecp}) in that limit. Not surprisingly: when 
$\Lambda_{\tt dark} \rightarrow 0$, the dark energy condensate never forms in the first place, 
and so there is nothing to decay.

As we noted above, in focusing on the bounce action (\ref{bounce}) we have effectively ignored gravitational 
effects on tunneling. This is justified when (\ref{bounce}) controls the nucleation rates of small bubbles, which are
not significantly influenced by the background geometry. In that case, we can simply go to a large local Minkowski
frame and resort to flat space calculations. Let us now explicitly show this is consistent 
for the bubbles which we are considering. If we compare the 
radius of a bubble at nucleation (\ref{radnuc}) to the Hubble length $1/H_0 \sim 10^{33} eV^{-1}$ in the late universe,
\be
r_0 H_0 \simeq \frac{\cal T}{\Delta V} H_0 \simeq \frac{10^{-9} eV^3}{10^{-12} eV^4} \times 
10^{-33} eV \sim 10^{-30}
\, .
\label{radiusbub}
\ee
So our bubbles are about a millimeter across when they nucleate. This means that they will not be longer  
than the Hubble length even if we go back in past to when the 
Hubble length was $10^{30}$ shorter - i.e. all the way to the $TeV$
scale. Since by all accounts dark chiral symmetry breaking happens much later, as per the discussion in Sec. 2, 
we can rely on the nucleation channel that yields the bounce action (\ref{bounce}) and ignore 
other, slower, decay channels, as well as any gravitational corrections in the semiclassical approximation. 

Note that in this case we only have the discharges of the dark CP-violating phase $\theta_{\tt dark}$ and dark
energy $V_{\tt dark} = \frac{1}{2} {\cal X}_{\tt dark} \bigl(\theta_{\tt dark}\bigr)^2$. The up-charging of
$\theta_{\tt dark}$ cannot happen spontaneously without gravity by positivity of the tension 
${\cal T}$. An up-charge would increase the dark energy density inside the membrane, 
which would need $r_0$ to be negative by Eq. (\ref{radnuc}), 
since ${\cal T} > 0$ and $\Delta V<0$. Without gravitational contributions 
such solutions starting from initial vacuum 
are ruled out. Of course, in the real universe where gravity is present (but weak, as observed) 
such processes could occur in some curved regions but are very rare. Since reverse processes are much faster,
such regions would evolve back into flat space eventually. 

In any case, the discharge would drive the universe toward Minkowski since by 
the charge equation (\ref{qcdcharge}) and the magnetic flux quantization  
(\ref{quantized}), $\theta_{\tt dark} =0$ belongs to the discharge sequence, $V_{\tt dark}(0) =0$ is the
energy minimum per Eq. (\ref{finpots}), and the nucleation rates are fast, per Eqs. (\ref{fastdec}), (\ref{numerika}). 
If the nucleation rates are sufficiently fast, this kind of dark energy would disappear after a long but finite time, and avoid the future horizon inside our lightcone.

\section{Cosmology of Discretely Decaying Dark Energy} 

As we noted in the discussion up to this point, the presence of dark QCD-like theory which goes strong 
around $\Lambda_{\tt dark} \sim 10^{-3} eV$ guarantees emergence of dark energy with 
\be
\rho_{\tt dark}  \simeq \Bigr( \Lambda_{\tt dark} \Bigr)^4 \sim 10^{-12} eV^4 \, ,
\label{derho}
\ee
to match the observation. In a universe which was shaped by inflation at early times, the dark sector
before chiral symmetry breaking is not completely cold due to dark particle production right near the
end of inflation. It's temperature then cannot be less than the Hubble parameter $H_*$ at the end of inflation,
because at least due to gravitational particle production which follows from universality of gravity.
For any inflationary model above the electroweak breaking scale, $H_* > TeV^2/\mpl \sim 10^{-3} eV$,
and so the dark sector is initially populated by relativistic gauge theory matter.

As the universe cools, so does the dark sector, whose temperature scales like the visible sector temperature.
Thus, at some point $T_{\tt dark} \la 10^{-3} eV$, dark chiral symmetry breaking occurs, and condensate forms
with a specific value of $\theta_{\tt dark}$. Since the universe emerged from inflation, all of it which 
belonged to the same causal domain before inflation will be in the same ground state, with the same latent value of 
$\theta_{\tt dark}$, which controls the dark energy density when the condensate forms. 
This is identical to what happens with the evolution of the adiabatic vacuum of a quantum field theory during and
after inflation, and is referred to as the "Quantum No-Hair Theorem" \cite{Kaloper:2018zgi}. As a result, the dark
energy will be initially homogeneous throughout the large region created by inflation. Given that 
$T_{\tt visible} > T_{\tt dark}$, this will occur well before the visible sector temperature drops to the current levels. 
In principle, this can happen as late as the visible sector recombination and/or radiation-matter equality,
as we explained in Sec. 2.

This dark energy ``permafrost" will eventually come to dominate over the visible sector, at redshifts 
$z \la {\cal O}(1)$, simply by cosmic expansion. As a result, the very late universe will begin to accelerate.
With our specification of parameters of the dark sector, in Sec 5, the accelerating epoch will last at 
least until the present time, since the decay rate of dark energy, controlled by the dark membrane 
nucleation rate in Eqs. (\ref{nucratecp}), (\ref{fastdec}) is slow enough to suppress the production of bubbles
of true vacuum $\theta_{\tt dark} = 0$ surrounded by membranes for at least 10 billion years (\ref{fastdec}).

Indeed, we can estimate how the decay rate (\ref{fastdec}) compares to the inverse Hubble 4-volume in
an earlier epoch. Let's suppose that $\zeta$ is chosen to saturate the condition (\ref{fastdec}). 
Clearly, we can scale this condition at will, since the prefactor and the bounce action controlling $\Gamma$ 
do not depend 
on the environment. Thus, at some previous epoch with $H_* > H_0$, 
\be
\frac{\Gamma}{H_*^4} \sim \frac{\Gamma}{H_0^4} \Bigl(\frac{H_0}{H_*}\Bigr)^4 \sim \Bigl(\frac{H_0}{H_*}\Bigr)^4\, ,
\label{fastdecscaling}
\ee
after choosing $\zeta$ to set ${\Gamma}/{H_0^4} \sim 1$. Then scaling the Hubble parameter using 
the redshift formula for the $\Lambda CDM$ late universe, $H(z) = \sqrt{0.7} H_0(1+0.43 (1+z)^3)^{1/2}$, we get
\be
\frac{\Gamma}{H_*^4}(z) \sim \frac{1}{0.49(1 + 0.43(1+z)^3)^2}\, ,
\label{fastdecscalingf}
\ee
which shows that already at the redshift of $z\sim 1$ the probability for a 
membrane to nucleate is reduced by about an order of magnitude. 

This has interesting implications for observational 
signature of our discrete dark energy:
\begin{itemize} 
\item First of all, we can 
infer that the decay rates for the membranes with tension and charge 
in the $10^{-3} eV$ range are ball park correct to trigger the dark energy meltdown about now. 
\item Secondly, this couldn't have started too soon, for the dark energy decays are 
local processes which occur by the formation of small 
bubbles, with a radius of about a millimeter, inside which there is no dark energy. The bubbles then 
expand at the speed of light taking over more and more of the universe. Therefore, any such small bubble 
would take about 10 	billion years to grow to the present Hubble length. Hence, e.g, if the 
nucleation probability was very low until about a $1/10$ of the current age of the universe, the 
large bubbles may be absent inside our Hubble volume. 
\item Third, this means
that dark energy may be getting very inhomogeneous currently at short scales, where it 
has begun evaporating away. If this process is currently underway, the nucleation 
rate is fast enough (\ref{nucratecp}) to yield the production of many bubbles now which can percolate and
completely remove the dark energy from 
the universe today \cite{Guth:1982pn,Turner:1992tz}. 
\end{itemize}
Note that as a consequence of the last point, at least our universe might not accelerate
forever, potentially escaping the conceptual quandary of future event horizons. 
In fact we can quantify this last point more precisely.

Near the end of Sec 4, we noted that
in contrast to the use of discrete discharges in addressing the strong CP problem 
\cite{Kaloper:2025wgn,Kaloper:2025upu}, in the case of 
dark energy the regime for tension and charge is less constrained. Nevertheless, there is still 
an interesting conjunction of numbers worth highlighting. As shown in \cite{Guth:1982pn}, 
the bubble nucleation rate in the Hubble units should satisfy 
$\Gamma/\hat H^4 \ga \frac{9n_c}{4\pi} \simeq 0.24$, to yield percolation and completely
stop vacuum energy driven inflation. Here $\hat H$ is the 
Hubble parameter during decays, and $n_c \sim 0.34$ is a numerical percolation factor. 
This along with the redshift dependence of $H$ in the $\Lambda CDM$ universe defines the 
`latest' value of $H$ which can be dissipated away. 

Since in our universe $\Omega_{DE} \simeq 0.7$
and $\Omega_{DM} \simeq 0.3$, the future asymptotic value of $H$, barring any other ``interventions", would be
$\hat H \simeq \sqrt{0.7} H_0 \sim 0.85 H_0$. A universe with this value of $\hat H$ will stop accelerating if
$\Gamma/\hat H^4 \ga 0.24$. Therefore $\Gamma/H_0^4$ in the present epoch 
must be $\Gamma/H_0^4 = (\Gamma/\hat H^4) (\hat H/H_0)^4 \ga 0.12$  
if the whole causal domain from our past is to transition to a state of zero vacuum energy in the future, and 
evade future event horizons. Since $\Gamma/H_0^4 \ga 0.12$ now, 
the probability for making bubbles sufficiently far in the past
so they can grow big to be easily observed is quite small. As we noted above, at $z\sim 1$ it would be
$\la 0.01$, and ever so smaller before. Hence having a currently accelerating universe with relatively few 
observational signals indicating the impending end to cosmic acceleration appears plausible. 
Note, that we can take a pragmatic stance also, ignore horizons 
and cosmic eschatology, and simply ask if a model that fits current observations might show signs of
changing cosmic acceleration rate. In that case
a smaller value of $\Gamma$ would work fine, too.

As the bubbles expand and grow, eventually they will encounter each other and collide. 
When they collide, the membranes would ignite and burn away by the production of dark sector particles, 
e.g. dark pions and glueballs. This follows since below the chiral symmetry breaking 
the theory includes the term ${\cal L}_{\tt dark} \ni \sqrt{\cal X} 
\theta_{\tt dark} \epsilon^{\mu\nu\lambda\sigma} {\cal F}_{\mu\nu\lambda\sigma} \sim 
\theta_{\tt dark} \partial_\mu K^\mu$, and so 
after a chiral transformation the Chern-Simons
current shifts on shell to $\partial K \rightarrow \partial K + f_\pi \partial^2 {\cal P}_{\tt dark} $, 
with ${\cal P}_{\tt dark} $ denoting the dark pion, and where $f_\pi \sim 10^{-3} eV$ 
is the dark pion decay constant. Combining these equations, we find 
${\cal L}_{\tt dark} \ni f_\pi  \theta_{\tt dark} \partial^2 {\cal P}_{\tt dark} $. 
Integrating it by parts and recalling that $\Delta  \theta \simeq 1$ 
on a membrane, $f_\pi \int d^4x \,  \theta_{\tt dark} \partial^2 {\cal P}_{\tt dark}  
\sim f_\pi \int_{R \times S^2} d^3 \xi \, n \cdot \partial {\cal P}_{\tt dark}$, where again $n$ is 
the outward normal to the membrane. Therefore, two membranes colliding with each other will  
produce dark pions, transferring their energy to the dark pion shower. 

Additionally, the colliding bubbles in the very late universe may also produce very long wavelength 
gravity waves. The mechanism is similar to the production of gravity waves during first order phase
transitions in the early universe \cite{Kosowsky:1992rz,Kamionkowski:1993fg}, but there are important differences. 
Here, the bubbles nucleate and collide very late, when the universe is already undergoing the late 
stage cosmic acceleration, rather than during radiation domination. This affects the estimates 
of the frequency and power of gravity waves produced by first
order phase transitions in comparison to the commonly used estimates for primordial gravity waves. 
Nevertheless, we can still eyeball the order of magnitude of these effects as required by the 
late universe cosmological consistency.

First of all, suppose a bubble forms, with an initial radius of about a millimeter, as we observed near the
end of Sec. 5. Initially, this bubble is small and has very little energy, but it is expanding at the speed of light. 
After a period $\tau$, which also serves as the lookback time to the event corresponding to the ``birth" of the bubble, 
it's size will be roughly $|\vec x| \sim \tau$, and its total energy, mostly being the
kinetic energy of the wall \cite{Coleman:1977py},  
\be
E_{\tt wall} \sim \frac{4\pi}{3} \Delta V |\vec x|^3 \sim \frac{4\pi}{3} \Delta V \tau^3 \, .
\label{walle}
\ee
The bubble ``birth" and growth occurs late, and so it can't be impeded by (dark) radiation. This 
simplifies the picture relative to the early universe processes 
\cite{Kosowsky:1992rz,Kamionkowski:1993fg}. If this energy were too large, which would happen when the bubble is 
comparable to the Hubble horizon size $\sim 1/H_0$, it could produce a large deformation
of the geometry and induce large anisotropies and inhomogeneities in the CMB. Thus a typical bubble shouldn't  
have grown too large yet, meaning that $\tau$ should be bounded from above. 

To get the idea of the bound, we follow the argument of \cite{Zeldovich:1974uw}, and require that
the total energy of the bubble is smaller than the total energy integrated over a Hubble volume,
$E_{\tt hubble} \simeq \rho_0/H_0^3$, by a factor of about $10^{-6}-10^{-7}$ to avoid distorting
CMB more than the concordance model. Since $\Delta V \simeq \rho_0 \simeq 10^{-12} eV^4$
for our model,  this yields roughly
\be
\tau H_0 \la 0.01 \, ,
\label{tau}
\ee
This means that the typical bubbles, if they have already formed, shouldn't be much larger than
about $100 MPc \sim 10^8$ lightyears. This corresponds to the redshift $z \sim 0.01$. 
The bubbles that might have nucleated before this redshift could have grown too much (if they haven't collided with 
one another and burned away); those which 
nucleated after might still be small enough and ``light" enough to 
not distort CMB too much. Therefore, it is clear that
there may also be the ``goldilocks" bubbles which nucleated at just the right time to 
leave a possibly detectable imprint, that might be just around the corner. 

To estimate gravity wave abundance from the bubble collisions, let us use
some order-of-magnitude estimates based on collisions of large shells, which are maximally off-center.
In this case, we roughly constrain the wavelength of the emitted gravity waves  
by the gravitational radius of the 
largest shells, $\lambda \ga {\cal R}_0$, which given the estimate (\ref{walle}) is about
\be
{\cal R}_0 \sim \frac{{E}_{\tt wall}}{\mpl^2} \sim 10 H_0^2 \tau^3 
\sim 10 (H_0 \tau)^3 H_0^{-1} \la 10^{-5}/H_0 \, .
\label{wavelength}
\ee
This corresponds to the frequencies $f_0 \la c/{\cal R}_0 \simeq 10^{-12} Hz$. 

Then to get an estimate on the upper
limit on the gravity wave power relative to the critical energy 
density of the universe, we start with the assumption 
that in each collision, an ${\cal O}(1)$ fraction of the shell's kinetic energy is converted to gravity 
waves\footnote{This is somewhat exaggerated, for the purpose of illustrating the potential for the 
signatures, and getting a definitive upper bound. Clearly, our numerical gravity wave 
estimates should be taken merely as guidelines rather than a firm codex.}. The
total energy density of such gravity waves would be, using (\ref{walle}) 
\be
\rho_{\tt GW} < {\cal N} \Delta V \tau^3 H_0^{3} \, ,   
\ee
where ${\cal N}$ is the number of events, and we have divided the total energy ${\cal N} E_{\tt wall}$ by the
Hubble volume $H_0^{-3}$. Then, $\Omega_{\tt GW} = \rho_{\tt GW}/\rho_0$, and again using $\Delta V \sim \rho_0$,
we find, using (\ref{tau}), and estimating the number of events as one per a volume of size $\tau^3$ inside a region 
of the universe of the scale of $(10 \tau)^3$, within which we see the bubbles nucleating (the regions farther
out being too young to contain bubbles),
\be
\Omega_{\tt GW} <  {\cal N}  \tau^3 H_0^{3} \sim 10^{-9} \, .  
\ee

Note that these gravity wave relics, such as they are, could currently be in an observational blindspot. 
First of all, they are very large wavelength, $\lambda \ga {\cal R}_0 \sim 100 kPc$, and low frequency,
$f_0 \la 10^{-12} Hz$. 
Further, the gravity wave amplitude is by necessity small, $\Omega_{\tt GW} < 10^{-9}$,
since otherwise the dark energy decay could disturb the CMB anisotropy too much. 
In comparison, the CMB observations can probe primordial B-mode imprints from gravity waves generated during inflation currently down to 
\cite{BICEP:2021xfz} 
\be
\Omega_{\tt GW} \sim (\frac{\delta \rho}{\rho})\Bigr|_{\tt T} \sim \sqrt{r} (\frac{\delta \rho}{\rho})\Bigr|_{\tt S} < 10^{-7} \, ,
\label{OMGW}
\ee
at scales which could go down to roughly 1\textdegree \, arc-length on the sky, that corresponds to wavelengths 
of about $\lambda \ga 2MPc$. The gravity waves which could be formed by our colliding bubbles are not
only shorter wavelength and lower power, but they cannot imprint in the CMB in the usual way since they appear
well after the surface of last scattering\footnote{We thank A. Westphal for a discussion of this point.}. 
However, the very recently proposed variation of the technique of using pulsar timing 
delays to search for gravity waves \cite{Hobbs:2009yy}
seems closer to being able to look for such ultralow 
frequency gravity waves \cite{DeRocco:2023qae}. Thus future explorations of 
pulsar timing delays might in fact provide a tool to explore our discretely evanescent dark energy.

\section{Summary} 

Dark energy models which are not just a cosmological constant almost invariably involve an ultralight 
scalar field, which rolls on a vey shallow potential along a very large distance in field space.
This requires fine-tuning and protecting 1) the magnitude of the potential 2) the curvature of the potential
and 3) the width of the field range. Further, the presence of an ultralight can also lead to new long range 
forces between other particles, and so on.

In contrast here we have presented a model for dark energy which does not involve any ultralight
scalar fields, nor does it require the variables controlling the appearance and 
eventual disappearance of dark energy to
vary over a huge range. Our model is a dark sector QCD-like theory, which is asymptotically free, and
where the scale of dark energy is obtained by dimensional transmutation. Due to the dark gluon and dark quark
content, the theory flows from the extreme UV, where it couples weakly, to the strong coupling regime 
and dark confinement in the far IR, 
which occurs around $\Lambda_{\tt dark} \sim 10^{-3} eV$. Thus this scale is 
technically natural and even natural, and the theory is UV-complete, at least as a 
quantum field theory. It could even
be unified with the visible sector around the GUT scale.

Once the theory enters strong coupling, chiral symmetry breaks via the formation of a condensate,
which also breaks dark sector CP. The condensate yields nonzero topological susceptibility and
CP breaking with the net dark vacuum angle, which is naturally ${\cal O}(1)$. Hence a locally constant
dark energy density $V_{\tt dark} \sim 10^{-12} eV^4$ naturally arises. However, the nontrivial topology of
the dark QCD gives rise to the emergent top form and membranes charged under it, 
which can nucleate by quantum effects. This provides relaxation channels for restoring dark CP and
evaporating away dark energy, which do not involve any ultralight fields or truly local degrees of freedom.
Instead, membrane nucleations relax the CP-breaking vacuum angle
toward zero, which modulo $2\pi$ is the only CP-invariant sector of the theory. If the nucleation rate
is fast enough, $\Gamma \ga H_0^4$, this will lead to complete disappearance of vacuum energy
in the entire universe. Such evolution could help alleviate some of 
the problems with cosmological horizons, which may not
arise in our future light cone.

Due to the relaxation mechanism via quantum tunnelling, the very late universe may eventually become 
quite inhomogeneous at late times, until the produced bubbles percolate. Further, bubble collisions might yield
very long wavelength gravity waves, that might be near the observational thresholds. We discussed the
cosmology of dark energy evolution from inflation onward, and sketched out
the late universe signatures. It would be interesting to analyze those in a more precise way.

\vskip1cm

{\bf Acknowledgments}: We thank G. D'Amico and especially A. Westphal 
for useful discussions. The research reported here 
was supported in part by the DOE Grant DE-SC0009999.


\begin{thebibliography}{99}

%\cite{DESI:2024mwx}
\bibitem{DESI:2024mwx}
A.~G.~Adame \textit{et al.} [DESI],
``DESI 2024 VI: cosmological constraints from the measurements of baryon acoustic oscillations,''
JCAP \textbf{02}, 021 (2025)
%doi:10.1088/1475-7516/2025/02/021
[arXiv:2404.03002 [astro-ph.CO]].
%896 citations counted in INSPIRE as of 25 May 2025

%\cite{Weinberg:1987dv}
\bibitem{Weinberg:1987dv}
S.~Weinberg,
``Anthropic Bound on the Cosmological Constant,''
Phys. Rev. Lett. \textbf{59}, 2607 (1987).
%doi:10.1103/PhysRevLett.59.2607
%1104 citations counted in INSPIRE as of 25 May 2025

%\cite{Martel:1997vi}
\bibitem{Martel:1997vi}
H.~Martel, P.~R.~Shapiro and S.~Weinberg,
``Likely values of the cosmological constant,''
Astrophys. J. \textbf{492}, 29 (1998)
%doi:10.1086/305016
[arXiv:astro-ph/9701099 [astro-ph]].
%273 citations counted in INSPIRE as of 25 May 2025

%\cite{Wetterich:1994bg}
\bibitem{Wetterich:1994bg}
C.~Wetterich,
``The Cosmon model for an asymptotically vanishing time dependent cosmological 'constant',''
Astron. Astrophys. \textbf{301}, 321-328 (1995)
[arXiv:hep-th/9408025 [hep-th]].
%1001 citations counted in INSPIRE as of 25 May 2025

%\cite{Zlatev:1998tr}
\bibitem{Zlatev:1998tr}
I.~Zlatev, L.~M.~Wang and P.~J.~Steinhardt,
``Quintessence, cosmic coincidence, and the cosmological constant,''
Phys. Rev. Lett. \textbf{82}, 896-899 (1999)
%doi:10.1103/PhysRevLett.82.896
[arXiv:astro-ph/9807002 [astro-ph]].
%2646 citations counted in INSPIRE as of 25 May 2025

%\cite{Weinberg:1988cp}
\bibitem{Weinberg:1988cp}
S.~Weinberg,
``The Cosmological Constant Problem,''
Rev. Mod. Phys. \textbf{61}, 1-23 (1989).
%doi:10.1103/RevModPhys.61.1
%6671 citations counted in INSPIRE as of 25 May 2025

%\cite{Polchinski:2006gy}
\bibitem{Polchinski:2006gy}
J.~Polchinski,
``The Cosmological Constant and the String Landscape,''
[arXiv:hep-th/0603249 [hep-th]].
%312 citations counted in INSPIRE as of 25 May 2025

%\cite{Arkani-Hamed:2006emk}
\bibitem{Arkani-Hamed:2006emk}
N.~Arkani-Hamed, L.~Motl, A.~Nicolis and C.~Vafa,
``The String landscape, black holes and gravity as the weakest force,''
JHEP \textbf{06}, 060 (2007)
%doi:10.1088/1126-6708/2007/06/060
[arXiv:hep-th/0601001 [hep-th]].
%1387 citations counted in INSPIRE as of 25 May 2025

%\cite{Frieman:1995pm}
\bibitem{Frieman:1995pm}
J.~A.~Frieman, C.~T.~Hill, A.~Stebbins and I.~Waga,
``Cosmology with ultralight pseudo Nambu-Goldstone bosons,''
Phys. Rev. Lett. \textbf{75}, 2077-2080 (1995)
%doi:10.1103/PhysRevLett.75.2077
[arXiv:astro-ph/9505060 [astro-ph]].
%903 citations counted in INSPIRE as of 25 May 2025

%\cite{Fukugita:1994hq}
\bibitem{Fukugita:1994hq}
M.~Fukugita and T.~Yanagida,
``Model for the cosmological constant,''
YITP-K-1098.
%6 citations counted in INSPIRE as of 25 May 2025

%\cite{Nomura:1999py}
\bibitem{Nomura:1999py}
Y.~Nomura, T.~Watari and T.~Yanagida,
``Mass generation for an ultralight axion,''
Phys. Rev. D \textbf{61}, 105007 (2000)
%doi:10.1103/PhysRevD.61.105007
[arXiv:hep-ph/9911324 [hep-ph]].
%17 citations counted in INSPIRE as of 25 May 2025 

%\cite{Armendariz-Picon:2000nqq}
\bibitem{Armendariz-Picon:2000nqq}
C.~Armendariz-Picon, V.~F.~Mukhanov and P.~J.~Steinhardt,
``A Dynamical solution to the problem of a small cosmological constant and late time cosmic acceleration,''
Phys. Rev. Lett. \textbf{85}, 4438-4441 (2000)
%doi:10.1103/PhysRevLett.85.4438
[arXiv:astro-ph/0004134 [astro-ph]].
%2008 citations counted in INSPIRE as of 25 May 2025


%\cite{Armendariz-Picon:2000ulo}
\bibitem{Armendariz-Picon:2000ulo}
C.~Armendariz-Picon, V.~F.~Mukhanov and P.~J.~Steinhardt,
``Essentials of k essence,''
Phys. Rev. D \textbf{63}, 103510 (2001)
%doi:10.1103/PhysRevD.63.103510
[arXiv:astro-ph/0006373 [astro-ph]].
%1911 citations counted in INSPIRE as of 25 May 2025

%\cite{Kim:2002tq}
\bibitem{Kim:2002tq}
J.~E.~Kim and H.~P.~Nilles,
``A Quintessential axion,''
Phys. Lett. B \textbf{553}, 1-6 (2003)
%doi:10.1016/S0370-2693(02)03148-9
[arXiv:hep-ph/0210402 [hep-ph]].
%130 citations counted in INSPIRE as of 25 May 2025

%\cite{Kaloper:2008qs}
\bibitem{Kaloper:2008qs}
N.~Kaloper and L.~Sorbo,
``Where in the String Landscape is Quintessence,''
Phys. Rev. D \textbf{79}, 043528 (2009)
%doi:10.1103/PhysRevD.79.043528
[arXiv:0810.5346 [hep-th]].
%94 citations counted in INSPIRE as of 25 May 2025

%\cite{DAmico:2018mnx}
\bibitem{DAmico:2018mnx}
G.~D'Amico, N.~Kaloper and A.~Lawrence,
``Strongly Coupled Quintessence,''
Phys. Rev. D \textbf{100}, no.10, 103504 (2019)
%doi:10.1103/PhysRevD.100.103504
[arXiv:1809.05109 [hep-th]].
%48 citations counted in INSPIRE as of 25 May 2025

%\cite{Hellerman:2001yi}
\bibitem{Hellerman:2001yi}
S.~Hellerman, N.~Kaloper and L.~Susskind,
``String theory and quintessence,''
JHEP \textbf{06}, 003 (2001)
%doi:10.1088/1126-6708/2001/06/003
[arXiv:hep-th/0104180 [hep-th]].
%351 citations counted in INSPIRE as of 25 May 2025

%\cite{Fischler:2001yj}
\bibitem{Fischler:2001yj}
W.~Fischler, A.~Kashani-Poor, R.~McNees and S.~Paban,
``The Acceleration of the universe, a challenge for string theory,''
JHEP \textbf{07}, 003 (2001)
%doi:10.1088/1126-6708/2001/07/003
[arXiv:hep-th/0104181 [hep-th]].
%316 citations counted in INSPIRE as of 25 May 2025


%\cite{Kaloper:2025wgn} 
\bibitem{Kaloper:2025wgn}
N.~Kaloper,
``An Alternative to Axion,''
[arXiv:2504.21078 [hep-ph]].
%1 citations counted in INSPIRE as of 25 May 2025

%\cite{Kaloper:2025upu}
\bibitem{Kaloper:2025upu}
N.~Kaloper,
``A Quantal Theory of Restoration of Strong CP Symmetry,''
[arXiv:2505.04690 [hep-ph]].
%1 citations counted in INSPIRE as of 25 May 2025

%\cite{Parker:1968mv}
\bibitem{Parker:1968mv}
L.~Parker,
%``Particle creation in expanding universes,''
Phys. Rev. Lett. \textbf{21}, 562-564 (1968). 
%doi:10.1103/PhysRevLett.21.562
%781 citations counted in INSPIRE as of 26 Mar 2024

%\cite{Bunch:1978yq} 
\bibitem{Bunch:1978yq}
T.~S.~Bunch and P.~C.~W.~Davies,
%``Quantum Field Theory in de Sitter Space: Renormalization by Point Splitting,''
Proc. Roy. Soc. Lond. A \textbf{360}, 117-134 (1978). 
%doi:10.1098/rspa.1978.0060
%1248 citations counted in INSPIRE as of 29 Mar 2024

%\cite{Linde:1990flp}
\bibitem{Linde:1990flp}
A.~D.~Linde, ``Particle physics and inflationary cosmology,''
Contemp. Concepts Phys. \textbf{5}, 1-362 (1990)
[arXiv:hep-th/0503203 [hep-th]].
%848 citations counted in INSPIRE as of 21 Mar 2024

 %\cite{Berezhiani:2000gw}
\bibitem{Berezhiani:2000gw}
Z.~Berezhiani, D.~Comelli and F.~L.~Villante,
``The Early mirror universe: Inflation, baryogenesis, nucleosynthesis and dark matter,''
Phys. Lett. B \textbf{503}, 362-375 (2001)
%doi:10.1016/S0370-2693(01)00217-9
[arXiv:hep-ph/0008105 [hep-ph]].
%312 citations counted in INSPIRE as of 02 Jun 2025
 
 %\cite{DAmico:2017lqj}
\bibitem{DAmico:2017lqj}
G.~D'Amico, P.~Panci, A.~Lupi, S.~Bovino and J.~Silk,
``Massive Black Holes from Dissipative Dark Matter,''
Mon. Not. Roy. Astron. Soc. \textbf{473}, no.1, 328-335 (2018)
%doi:10.1093/mnras/stx2419
[arXiv:1707.03419 [astro-ph.CO]].
%54 citations counted in INSPIRE as of 02 Jun 2025

%\cite{BICEP:2021xfz}
\bibitem{BICEP:2021xfz}
P.~A.~R.~Ade \textit{et al.} [BICEP and Keck],
``Improved Constraints on Primordial Gravitational Waves using 
Planck, WMAP, and BICEP/Keck Observations through the 2018 Observing Season,''
Phys. Rev. Lett. \textbf{127}, no.15, 151301 (2021)
%doi:10.1103/PhysRevLett.127.151301
[arXiv:2110.00483 [astro-ph.CO]].
%997 citations counted in INSPIRE as of 25 May 2025

%\cite{Georgi:1984zwz}
\bibitem{Georgi:1984zwz}
H.~Georgi,
``Weak Interactions and Modern Particle Theory,''
%30 citations counted in INSPIRE as of 26 May 2025
 Benjamin-Cummings Pub Co, 1984. 
 
 
%\cite{Luscher:1978rn}
\bibitem{Luscher:1978rn}
M.~L\"uscher, ``The Secret Long Range Force in Quantum Field Theories With Instantons,''
Phys. Lett. B \textbf{78}, 465-467 (1978).
%doi:10.1016/0370-2693(78)90487-2
%158 citations counted in INSPIRE as of 12 Aug 2024
 
 
%\cite{Dvali:2005an}
\bibitem{Dvali:2005an}
G.~Dvali, ``Three-form gauging of axion symmetries and gravity,''
[arXiv:hep-th/0507215 [hep-th]].
%143 citations counted in INSPIRE as of 19 Apr 2023

%\cite{Dvali:2005zk}
\bibitem{Dvali:2005zk}
G.~Dvali,
``A Vacuum accumulation solution to the strong CP problem,''
Phys. Rev. D \textbf{74}, 025019 (2006)
%doi:10.1103/PhysRevD.74.025019
[arXiv:hep-th/0510053 [hep-th]].
%44 citations counted in INSPIRE as of 12 Aug 2024

%\cite{Aurilia:1978qs}
\bibitem{Aurilia:1978qs}
A.~Aurilia, D.~Christodoulou and F.~Legovini,
``A Classical Interpretation of the Bag Model for Hadrons,''
Phys. Lett. B \textbf{73}, 429-432 (1978).
%doi:10.1016/0370-2693(78)90757-8
%39 citations counted in INSPIRE as of 29 May 2025

%\cite{Aurilia:1980xj}
\bibitem{Aurilia:1980xj}
A.~Aurilia, H.~Nicolai and P.~K.~Townsend,
``Hidden Constants: The Theta Parameter of 
QCD and the Cosmological Constant of N=8 Supergravity,''
Nucl. Phys. B \textbf{176}, 509-522 (1980). 
%doi:10.1016/0550-3213(80)90466-6
%222 citations counted in INSPIRE as of 21 Nov 2021

%\cite{Duff:1980qv}
\bibitem{Duff:1980qv}
M.~J.~Duff and P.~van Nieuwenhuizen,
``Quantum Inequivalence of Different Field Representations,''
Phys. Lett. B \textbf{94}, 179-182 (1980). 
%doi:10.1016/0370-2693(80)90852-7
%304 citations counted in INSPIRE as of 01 May 2025

%\cite{Aurilia:1980jz}
\bibitem{Aurilia:1980jz}
A.~Aurilia, Y.~Takahashi and P.~K.~Townsend,
``The U(1) Problem and the Higgs Mechanism in Two-dimensions and Four-dimensions,''
Phys. Lett. B \textbf{95}, 265-268 (1980).
%doi:10.1016/0370-2693(80)90484-0
%79 citations counted in INSPIRE as of 18 Aug 2024

%\cite{Vafa:1984xg}
\bibitem{Vafa:1984xg}
C.~Vafa and E.~Witten,
``Parity Conservation in QCD,''
Phys. Rev. Lett. \textbf{53}, 535 (1984). %
%doi:10.1103/PhysRevLett.53.535
%537 citations counted in INSPIRE as of 12 Aug 2024


%\cite{Gabadadze:1997kj}
\bibitem{Gabadadze:1997kj}
G.~Gabadadze,
``Modeling the glueball spectrum by a closed bosonic membrane,''
Phys. Rev. D \textbf{58}, 094015 (1998)
%doi:10.1103/PhysRevD.58.094015
[arXiv:hep-ph/9710402 [hep-ph]].
%34 citations counted in INSPIRE as of 20 Aug 2024

%\cite{Gabadadze:2002ff}
\bibitem{Gabadadze:2002ff}
G.~Gabadadze and M.~Shifman,
``QCD vacuum and axions: What's happening?,''
Int. J. Mod. Phys. A \textbf{17}, 3689-3728 (2002)
%doi:10.1142/S0217751X02011357
[arXiv:hep-ph/0206123 [hep-ph]].
%70 citations counted in INSPIRE as of 20 Aug 2024

%\cite{Aurilia:1978dw}
\bibitem{Aurilia:1978dw}
A.~Aurilia, ``The Problem of Confinement: From Two-dimensions to Four-dimensions,''
Phys. Lett. B \textbf{81}, 203-206 (1979). 
%doi:10.1016/0370-2693(79)90524-0
%45 citations counted in INSPIRE as of 12 Aug 2024


%\cite{Gnadig:1976pn}
\bibitem{Gnadig:1976pn}
P.~Gnadig, P.~Hasenfratz, J.~Kuti and A.~S.~Szalay,
``The Quark Bag Model with Surface Tension,''
Phys. Lett. B \textbf{64}, 62-66 (1976).
%doi:10.1016/0370-2693(76)90358-0
%44 citations counted in INSPIRE as of 13 Aug 2024


%\cite{Brown:1987dd}
\bibitem{Brown:1987dd}
J.~D.~Brown and C.~Teitelboim,
``Dynamical Neutralization of the Cosmological Constant,''
Phys. Lett. B \textbf{195}, 177-182 (1987). 
%doi:10.1016/0370-2693(87)91190-7
%268 citations counted in INSPIRE as of 14 Jan 2022

%\cite{Brown:1988kg}
\bibitem{Brown:1988kg}
J.~D.~Brown and C.~Teitelboim,
``Neutralization of the Cosmological Constant by Membrane Creation,''
Nucl. Phys. B \textbf{297}, 787-836 (1988). 
%doi:10.1016/0550-3213(88)90559-7
%353 citations counted in INSPIRE as of 21 Nov 2021


%\cite{Kaloper:2022oqv}
\bibitem{Kaloper:2022oqv}
N.~Kaloper,
``Hidden variables of gravity and geometry and the cosmological constant problem,''
Phys. Rev. D \textbf{106}, no.6, 065009 (2022)
%doi:10.1103/PhysRevD.106.065009
[arXiv:2202.06977 [hep-th]].
%9 citations counted in INSPIRE as of 19 Apr 2023

%\cite{Kaloper:2022utc} 
\bibitem{Kaloper:2022utc}
N.~Kaloper,
``Pancosmic Relativity and Nature's Hierarchies,''
Phys. Rev. D \textbf{106}, no.4, 044023 (2022)
%doi:10.1103/PhysRevD.106.044023
[arXiv:2202.08860 [hep-th]].
%3 citations counted in INSPIRE as of 19 Apr 2023

%\cite{Schwinger:1951nm} 
\bibitem{Schwinger:1951nm}
J.~S.~Schwinger,
``On gauge invariance and vacuum polarization,''
Phys. Rev. \textbf{82}, 664-679 (1951).
%doi:10.1103/PhysRev.82.664
%5938 citations counted in INSPIRE as of 30 Jul 2023


%\cite{Shifman:1998if}
\bibitem{Shifman:1998if}
M.~A.~Shifman,
``Domain walls and decay rate of the excited vacua in the large N Yang-Mills theory,''
Phys. Rev. D \textbf{59}, 021501 (1999)
%doi:10.1103/PhysRevD.59.021501
[arXiv:hep-th/9809184 [hep-th]].
%51 citations counted in INSPIRE as of 23 Mar 2025

%\cite{Forbes:2000et}
\bibitem{Forbes:2000et}
M.~M.~Forbes and A.~R.~Zhitnitsky, ``Domain walls in QCD,''
JHEP \textbf{10}, 013 (2001)
%doi:10.1088/1126-6708/2001/10/013
[arXiv:hep-ph/0008315 [hep-ph]].
%49 citations counted in INSPIRE as of 14 Apr 2025

%\cite{Dvali:1998ms}
\bibitem{Dvali:1998ms}
G.~R.~Dvali and Z.~Kakushadze,
``Large N domain walls as D-branes for N=1 QCD string,''
Nucl. Phys. B \textbf{537}, 297-316 (1999)
%doi:10.1016/S0550-3213(98)00683-X
[arXiv:hep-th/9807140 [hep-th]].
%34 citations counted in INSPIRE as of 24 Apr 2025

%\cite{Dubovsky:2011tu}
\bibitem{Dubovsky:2011tu}
S.~Dubovsky, A.~Lawrence and M.~M.~Roberts,
``Axion monodromy in a model of holographic gluodynamics,''
JHEP \textbf{02}, 053 (2012)
%doi:10.1007/JHEP02(2012)053
[arXiv:1105.3740 [hep-th]].
%67 citations counted in INSPIRE as of 27 Mar 2025


%\cite{Coleman:1977py}
\bibitem{Coleman:1977py}
S.~R.~Coleman,
``The Fate of the False Vacuum. 1. Semiclassical Theory,''
Phys. Rev. D \textbf{15}, 2929-2936 (1977)
[erratum: Phys. Rev. D \textbf{16}, 1248 (1977)].
%doi:10.1103/PhysRevD.16.1248
%2626 citations counted in INSPIRE as of 12 Aug 2024

%\cite{Callan:1977pt}
\bibitem{Callan:1977pt}
C.~G.~Callan, Jr. and S.~R.~Coleman,
``The Fate of the False Vacuum. 2. First Quantum Corrections,''
Phys. Rev. D \textbf{16}, 1762-1768 (1977). 
%doi:10.1103/PhysRevD.16.1762
%1765 citations counted in INSPIRE as of 05 Apr 2025

%\cite{Garriga:1993fh}
\bibitem{Garriga:1993fh}
J.~Garriga,
``Nucleation rates in flat and curved space,''
Phys. Rev. D \textbf{49}, 6327-6342 (1994)
%doi:10.1103/PhysRevD.49.6327
[arXiv:hep-ph/9308280 [hep-ph]].
%98 citations counted in INSPIRE as of 12 Aug 2024


%\cite{Guth:1982pn}
\bibitem{Guth:1982pn}
A.~H.~Guth and E.~J.~Weinberg,
``Could the Universe Have Recovered from a Slow First Order Phase Transition?,''
Nucl. Phys. B \textbf{212}, 321-364 (1983).
%doi:10.1016/0550-3213(83)90307-3
%662 citations counted in INSPIRE as of 14 May 2025

%\cite{Turner:1992tz}
\bibitem{Turner:1992tz}
M.~S.~Turner, E.~J.~Weinberg and L.~M.~Widrow,
``Bubble nucleation in first order inflation and other cosmological phase transitions,''
Phys. Rev. D \textbf{46}, 2384-2403 (1992).
%doi:10.1103/PhysRevD.46.2384
%284 citations counted in INSPIRE as of 14 May 2025

%\cite{Kaloper:2018zgi}
\bibitem{Kaloper:2018zgi}
N.~Kaloper and J.~Scargill,
``Quantum Cosmic No-Hair Theorem and Inflation,''
Phys. Rev. D \textbf{99}, no.10, 103514 (2019)
%doi:10.1103/PhysRevD.99.103514
[arXiv:1802.09554 [hep-th]].
%21 citations counted in INSPIRE as of 29 May 2025

%\cite{Kosowsky:1992rz}
\bibitem{Kosowsky:1992rz} 
A.~Kosowsky, M.~S.~Turner and R.~Watkins,
``Gravitational waves from first order cosmological phase transitions,''
Phys. Rev. Lett. \textbf{69}, 2026-2029 (1992). 
%doi:10.1103/PhysRevLett.69.2026
%485 citations counted in INSPIRE as of 12 Apr 2025

%\cite{Kamionkowski:1993fg}
\bibitem{Kamionkowski:1993fg}
M.~Kamionkowski, A.~Kosowsky and M.~S.~Turner,
``Gravitational radiation from first order phase transitions,''
Phys. Rev. D \textbf{49}, 2837-2851 (1994)
%doi:10.1103/PhysRevD.49.2837
[arXiv:astro-ph/9310044 [astro-ph]].
%883 citations counted in INSPIRE as of 24 Mar 2025

%\cite{Zeldovich:1974uw}
\bibitem{Zeldovich:1974uw}
Y.~B.~Zeldovich, I.~Y.~Kobzarev and L.~B.~Okun,
``Cosmological Consequences of the Spontaneous Breakdown of Discrete Symmetry,''
Zh. Eksp. Teor. Fiz. \textbf{67}, 3-11 (1974). 
%SLAC-TRANS-0165.
%937 citations counted in INSPIRE as of 10 Sep 2023

%\cite{Hobbs:2009yy}
\bibitem{Hobbs:2009yy}
G.~Hobbs, A.~Archibald, Z.~Arzoumanian, D.~Backer, M.~Bailes, N.~D.~R.~Bhat, M.~Burgay, S.~Burke-Spolaor, D.~Champion and I.~Cognard, \textit{et al.}
``The international pulsar timing array project: using pulsars as a gravitational wave detector,''
Class. Quant. Grav. \textbf{27}, 084013 (2010)
%doi:10.1088/0264-9381/27/8/084013
[arXiv:0911.5206 [astro-ph.SR]].
%630 citations counted in INSPIRE as of 02 Jun 2025

%\cite{DeRocco:2023qae}
\bibitem{DeRocco:2023qae}
W.~DeRocco and J.~A.~Dror,
``Searching for stochastic gravitational waves below a nanohertz,''
Phys. Rev. D \textbf{108}, no.10, 103011 (2023)
%doi:10.1103/PhysRevD.108.103011
[arXiv:2304.13042 [astro-ph.HE]].
%23 citations counted in INSPIRE as of 31 May 2025

\end{thebibliography}
\end{document}